\documentclass[aps,preprint,prd,nofootinbib]{revtex4}

\usepackage{amsmath}
\usepackage{graphicx}
\usepackage{xcolor}
\RequirePackage[colorlinks,citecolor=blue,urlcolor=blue,hyperfootnotes=false]{hyperref}
\usepackage{subcaption}

\newcommand{\fpbh}{f_{\text{PBH}}}
\newcommand{\mpbh}{M_{\text{PBH}}}
\newcommand{\rcl}{r_{\text{cl}}}

\begin{document}

\title{Constraining supermassive primordial black hole clustering with the angular auto-correlation of $z\simeq 6$ quasars}

\author{Zhan-He Wang$^{1,2}$\footnote{\href{wangzhanhe19@mails.ucas.ac.cn}{wangzhanhe19@mails.ucas.ac.cn}}}
\author{Hai-Long Huang$^{1,2}$\footnote{\href{huanghailong18@mails.ucas.ac.cn}{huanghailong18@mails.ucas.ac.cn}}}
\author{Yun-Song Piao$^{1,2,3,4}$\footnote{\href{yspiao@ucas.ac.cn}{yspiao@ucas.ac.cn}}}

\affiliation{$^1$ School of Fundamental Physics and Mathematical
    Sciences, Hangzhou Institute for Advanced Study, UCAS, Hangzhou
    310024, China}

\affiliation{$^2$ School of Physical Sciences, University of
Chinese Academy of Sciences, Beijing 100049, China}

\affiliation{$^3$ International Center for Theoretical Physics
    Asia-Pacific, Beijing/Hangzhou, China}

\affiliation{$^4$ Institute of Theoretical Physics, Chinese
    Academy of Sciences, P.O. Box 2735, Beijing 100190, China}



\begin{abstract}
High-redshift quasars provide a direct probe of the origin and
environment of the earliest supermassive black holes. We use their
angular auto-correlation function at $z\simeq 6$ to test scenarios
in which supermassive primordial black holes (SMPBHs) are
associated with the observed quasar population. The evolved PBH
correlation functions, for both Poisson fluctuations and initial
PBH clustering, are projected over the quasar redshift window and
compared with the measured angular correlation function using
Markov chain Monte Carlo inference. It is observed that for the
Poisson model, the posterior favors a small abundance, $\fpbh\sim
10^{-3}$, and a supermassive effective mass scale, $\mpbh\sim
10^{12}M_\odot$, interpreted here as a scale controlling quasar
host-halo formation and clustering, and for the initially
clustered model, the data prefer an effective clustering amplitude
$\xi_{\rm eff}\simeq 2.1$ and a top-hat boundary scale $\rcl\simeq
76\,{\rm Mpc}$, corresponding to weak relative contraction of PBH
pairs in comoving coordinates.
\end{abstract}

\maketitle


\section{Introduction}

Wide-field optical and near-infrared surveys, including the Sloan
Digital Sky Survey (SDSS)
\cite{Fan:2001ff,Fan:2004ny,Jiang_2007,Jiang:2016eir}, Subaru
High-$z$ Exploration of Low-Luminosity Quasars (SHELLQs)
\cite{Matsuoka_2017,Matsuoka:2019zsd,Matsuoka_2021}, the
Canada--France--Hawaii Telescope Legacy Survey (CFHTLS)
\cite{Willott:2007tx,Willott:2010xf}, the Panoramic Survey
Telescope and Rapid Response System 1 (PanSTARRS1)
\cite{Banados_2016}, and the United Kingdom Infrared Telescope
Infrared Deep Sky Survey (UKIDSS) \cite{Mortlock:2011va}, have
discovered a large population of luminous quasars at $z\gtrsim 6$
\cite{Fan:2022TOIGM}. These observations have established that
supermassive black holes (SMBHs) with masses of order
$10^8$--$10^{10}M_\odot$ were already in place within the first
billion years of cosmic history
\cite{Wu:2015xaa,Onoue:2019glq,Farina:2022wub}. These objects
probe the formation of the earliest massive black holes and their
host environments, while also posing a long-standing growth
problem. If the seeds are from ordinary stellar remnants, the
available cosmic time is short and efficient, sustained accretion
is required. Heavy-seed scenarios, such as direct collapse or
dense stellar systems, can alleviate this tension but depend
sensitively on the local gas, radiation, and feedback environment
\cite{Volonteri:2021sfo,Inayoshi:2019fun}.

Primordial black holes (PBHs) provide a qualitatively different
possibility \cite{Hawking:1971ei,Carr:1974nx,Zeldovich:1967lct}.
In corresponding scenarios, PBHs formed in the very early Universe
from large primordial density fluctuations, their masses are not
tied to stellar evolution and may span a wide range, with
implications for dark matter and massive black hole formation
\cite{Carr:2016drx,Chapline:1975ojl,Meszaros:1975ef,Carr:2020xqk,Carr:2023tpt,Nakama:2016kfq,Huang:2025gwpop,Jiang:2025gwfast}.
Supermassive PBHs (SMPBHs) have therefore been considered as
possible progenitors of the SMBHs powering high-redshift quasars,
or more generally as early compact objects that affect the
formation of massive halos
\cite{Carr:2023tpt,Nakama:2016kfq,Huang:2024ghz,Huang:2024lrd,Li:2025uvlf}.\footnote{Related
early-universe settings, such as
vacuum decay and AdS phases, have also been discussed in
connection with large inhomogeneities, wormhole and
gravitational-wave signals
\cite{Lin:2021vwc,Lin:2022ygd,Li:2020cjj,Ye:2020btb,Jiang:2021bab,Ye:2021iwa,Wang:2024hks,Wang:2024dka,Lan:2024wormhole,Peng:2026deepinflation,Lai:2026echo}.}
This possibility is constrained by the required enhancement of the
primordial power spectrum, by CMB spectral distortions, and by
other cosmological and astrophysical bounds
\cite{Fixsen:1996nj,Mather:1993ij,Zeldovich:1969ff,Chluba:2011hw,Kohri:2014lza,Gow:2020bzo,Nakama:2017xvq}.
In particular, producing SMPBHs through enhanced curvature
perturbations can be in tension with $\mu$-distortion constraints,
although non-Gaussianity and broad power spectra may modify the
allowed parameter space
\cite{Sharma:2024cpl,Byrnes:2024yrp,Pritchard:2025ywt,Wang:2025muDistortion}.

In addition to their mass and abundance, the spatial distribution of PBHs
directly affects clustering observables. Even if PBHs are initially distributed
randomly, their discreteness generates a Poisson isocurvature component whose
nonlinear evolution can enhance small-scale clustering \cite{Luca_2020}. Such
Poisson initial conditions provide a minimal prediction, but PBH formation need
not produce a purely Poisson distribution. Non-Gaussian primordial fluctuations
and model-dependent mechanisms can generate beyond-Poisson correlations already
at formation
\cite{Tada:2015noa,Young:2013oia,Franciolini:2018vbk,Suyama:2019cst}. In
particular, SMPBHs produced from inflationary bubbles or related multistream
mechanisms can naturally exhibit initial clustering, which can enhance their
binary merger rate and affect the subsequent evolution of the cluster
\cite{Huang:2023chx,Huang:2023klk,Huang_2024,Hai-LongHuang:2023atg}. Initially
clustered PBH populations have been studied in connection with dark matter
constraints and early structure formation \cite{DeLuca_2022,Huang_2024}, but
the subsequent evolution of such initial clustering and its impact on quasar
angular clustering remain less explored.

In this work we study the angular auto-correlation function of $z\simeq 6$
quasars as a probe of PBH clustering. We consider two classes of initial
conditions. The first is the Poisson case, for which the PBH discreteness
determines the initial power spectrum and the correlation function evolves
through linear, quasi-linear, and nonlinear regimes. The second is an initially
clustered PBH population, for which we introduce a phenomenological evolution
model based on a spherical top-hat correlation profile \cite{DeLuca_2022,Huang_2024}
and pair conservation.
The evolved three-dimensional PBH correlation functions are projected over the
observed quasar redshift window and compared with the measured angular
correlation function, following the general strategy of using high-redshift
quasar clustering to test SMPBH scenarios
\cite{Shinohara:2021psq,Shinohara:2023wjd,Arita:2023xsp}.

The paper is organized as follows. Sec.~\ref{sec:poisson-evolution} reviews the
evolution from Poisson initial conditions and derives the corresponding
real-space correlation function. Sec.~\ref{sec:cluster-evolution} constructs a
phenomenological model for initially clustered PBH systems, while
Sec.~\ref{sec:observational-analysis} describes the quasar sample, random
catalog, survey mask, and angular auto-correlation estimator. In
Sec.~\ref{sec:results}, we compare the observed angular correlation function
with the Poisson and initially clustered PBH models using Markov chain Monte
Carlo (MCMC) inference. Sec.~\ref{sec:conclusions} summarizes the implications
and possible extensions.

\section{Evolution from Poisson initial conditions}
\label{sec:poisson-evolution}

In this section, we consider the evolution from Poisson initial conditions \cite{Luca_2020}. We first
introduce the PBH two-point correlation function as follows:

\begin{equation}
    \left\langle
    \frac{\delta\rho_{\rm PBH}(\vec{x},z)}{\bar{\rho}_{\rm PBH}}
    \frac{\delta\rho_{\rm PBH}(0,z)}{\bar{\rho}_{\rm PBH}}
    \right\rangle
    =
    \frac{1}{\bar{n}_{\rm PBH}}\delta_D(\vec{x})
    + \xi(\vec{x},z).
\end{equation}

Here $\delta_D(\vec{x})$ denotes the Dirac delta function, and the first term is the
Poisson contribution from the discreteness of PBHs. The second term
$\xi(\vec{x},z)$ describes the excess spatial correlation beyond the Poisson
component, which is absent for the Poisson initial conditions considered in this
section.

The mean comoving number density of PBHs is
\begin{equation}
    \bar{n}_{\rm PBH} \simeq
    3.2\,\fpbh
    \left( \frac{20\,M_{\odot}/h}{\mpbh} \right)
    (h/{\rm kpc})^3 ,
\end{equation}
where $\fpbh$ is the PBH fraction in dark matter and $\mpbh$ denotes the
characteristic PBH mass entering the discreteness scale of the model.

The evolution of the PBH power spectrum proceeds through three regimes: the
linear, quasi-linear, and nonlinear regimes.
The corresponding dimensionless power spectrum can be approximated by
\begin{equation}
    \label{eq:pbh-dimensionless-power-spectrum}
    \Delta^2_{\rm PBH}(k,z) \simeq
    \begin{cases}
        \left( 1+\dfrac{3}{2}\fpbh\dfrac{1+z_{\rm eq}}{1+z} \right)^2
        \Delta_i^2(k),
        & k < k_{\rm L-QL}(z), \\[1.2ex]
        \left( \dfrac{k}{k_{\rm L-QL}(z)} \right)^{9/4},
        & k_{\rm L-QL}(z) \leq k < k_{\rm QL-NL}(z), \\[1.2ex]
        200\left( \dfrac{k}{k_{\rm QL-NL}(z)} \right)^{9/5},
        & k \geq k_{\rm QL-NL}(z).
    \end{cases}
\end{equation}
The transition scales are given by
\begin{align}
    k_{\rm L-QL}(z)
    &\simeq
    \frac{4}{\fpbh^{1/3}}
    \left( \frac{20\,M_{\odot}/h}{\mpbh} \right)^{1/3}
    \left[
        1+26\fpbh\left( \frac{100}{1+z} \right)
    \right]^{-2/3}
    h/{\rm kpc}, \label{eq:k-l-ql} \\
    k_{\rm QL-NL}(z)
    &\simeq
    \frac{42}{\fpbh^{1/3}}
    \left( \frac{20\,M_{\odot}/h}{\mpbh} \right)^{1/3}
    \left[
        1+26\fpbh\left( \frac{100}{1+z} \right)
    \right]^{-2/3}
    h/{\rm kpc}. \label{eq:k-ql-nl}
\end{align}
Now we consider the relationship between $\xi(x,z)$ and $\Delta^2_{\rm PBH}(k,z)$.
We define the volume averaged correlation function as follows:
\begin{align}
    \bar{\xi}(R,z)
    &=
    \frac{3}{4\pi R^3}
    \int_0^R ds\,4\pi s^2\xi(s,z),
\end{align}
where
\begin{align}
    \xi(\vec{x},z)
    &\simeq
    \int \frac{dk}{k}\,
    e^{i\vec{k}\cdot\vec{x}}\Delta^2(k,z).
\end{align}
Therefore, the dimensionless power spectrum can be estimated as \cite{Luca_2020}
\begin{equation}
    \Delta^2(k,z) \simeq \bar{\xi}(1/k,z).
\end{equation}
With Eq.~\eqref{eq:pbh-dimensionless-power-spectrum}, we can approximately obtain the
corresponding evolution of the correlation function.
\begin{equation}
    \label{eq:poisson-correlation-evolution}
    \xi(x,z) \simeq
    \begin{cases}
        0,
        & x > x_{L-QL}(z), \\[1ex]
        \left(\dfrac{x_{L-QL}(z)}{x}\right)^{9/4}
        \sim \left( \dfrac{1}{1+z} \right)^{3/2},
        & x_{QL-NL}(z) \leq x < x_{L-QL}(z), \\[1ex]
        200\left(\dfrac{x_{QL-NL}(z)}{x}\right)^{9/5}
        \sim \left( \dfrac{1}{1+z} \right)^{6/5},
        & 0 < x \leq x_{QL-NL}(z).
    \end{cases}
\end{equation}
The transition lengths are the inverse of the transition wavenumbers in
Eqs.~\eqref{eq:k-l-ql} and \eqref{eq:k-ql-nl},
\begin{align}
    x_{L-QL}(z)
    &\simeq
    \frac{\fpbh^{1/3}}{4}
    \left( \frac{\mpbh}{20\,M_{\odot}/h} \right)^{1/3}
    \left[
        1+26\fpbh\left( \frac{100}{1+z} \right)
    \right]^{2/3}
    {\rm kpc}/h, \\
    x_{QL-NL}(z)
    &\simeq
    \frac{\fpbh^{1/3}}{42}
    \left( \frac{\mpbh}{20\,M_{\odot}/h} \right)^{1/3}
    \left[
        1+26\fpbh\left( \frac{100}{1+z} \right)
    \right]^{2/3}
    {\rm kpc}/h.
\end{align}

\section{Cluster evolution with initially clustered systems}
\label{sec:cluster-evolution}

For initially clustered PBHs, we assume the following initial
correlation function \cite{DeLuca_2022,Huang_2024}

\begin{equation}
    \label{eq:initial-cluster-correlation}
    \xi(x) =
    \begin{cases}
      \xi_0 & \text{if } ~ x \leq r_0 \\
       0 & \text{otherwise}
\end{cases}
\end{equation}

This top-hat profile provides a minimal phenomenological description of an
initially clustered PBH population. The parameter $\xi_0$ characterizes the
initial clustering amplitude, while $r_0$ denotes the comoving size of the
clustered region. The PBH density contrast is essentially frozen during the
radiation-dominated era, and the subsequent clustering evolution starts only
after matter-radiation equality. We therefore identify the onset redshift of the
phenomenological evolution with $z_{\rm eq}$, namely
$z_{\rm eq}\simeq 3400$ in the numerical analysis.
In what follows we track how this initially localized excess probability evolves
under the relative motion of PBH pairs.

We model the subsequent evolution by preserving the top-hat form of the
correlation function, while allowing both its amplitude and characteristic
radius to evolve. Thus we write
\begin{equation}
    \label{eq:cluster-evolution-ansatz}
    \xi(x,z) =
    \begin{cases}
      \xi_0 D(z)& \text{if } ~ x \leq \rcl \\
       0 & \text{otherwise}
\end{cases}
\end{equation}
where $D(z)$ encodes the growth of the clustered component. The evolution of
$D(z)$ is constrained by pair conservation. For an isotropic distribution, the
pair conservation equation gives \cite{peebles2020large}
\begin{equation}
    \label{eq:pair-conservation}
    \frac{\langle v_{12}(x,a) \rangle}{H(a)ax}=-\frac{1}{1+\xi(x,a)}\frac{a}{x^3}\frac{\partial}{\partial a} \int_{0}^{x} \xi(y,a)y^2dy
\end{equation}
where $\langle v_{12}\rangle$ is the mean pairwise peculiar velocity. To close
the system, we parametrize the clustering motion by interpolating between two
limiting cases: fixed comoving separations and stable physical separations
within the cluster,
\begin{equation}
    \label{eq:pairwise-velocity-ansatz}
    \langle v_{12}(x,a) \rangle = -\lambda H(a)ax, \qquad 0\le \lambda \le 1
\end{equation}
Here $\lambda=0$ corresponds to no contraction in comoving coordinates, whereas
$\lambda=1$ corresponds to stable clustering with approximately fixed physical
separations.

In the same phenomenological description, the characteristic cluster radius
shrinks in comoving coordinates after the onset of nonlinear evolution,
$\rcl = \left(\frac{a_{\rm eq}}{a}\right)^\lambda r_0$.
Substituting the top-hat ansatz in Eq.~\eqref{eq:cluster-evolution-ansatz} and
the velocity prescription in Eq.~\eqref{eq:pairwise-velocity-ansatz} into the
pair-conservation equation \eqref{eq:pair-conservation}, and fixing the
normalization with the initial condition \eqref{eq:initial-cluster-correlation},
we obtain
\begin{equation}
    \xi (x,z) =
\begin{cases}
      (1+\xi_0)\left( \frac{1+z_{\rm eq}}{1+z}\right)^{3\lambda}- 1  & \text{if } ~ x \leq \rcl \\
       0 & \text{otherwise}
\end{cases}
\end{equation}

\section{Quasar sampling and observational analysis}
\label{sec:observational-analysis}

The theoretical predictions obtained in the previous sections are projected
quantities on the sky and should therefore be confronted with the observed
angular distribution of high-redshift quasars. In this section we describe the
observational analysis pipeline used to construct the quasar angular
auto-correlation function. The procedure follows the standard pair-count
approach: we first define the quasar sample and its angular selection function,
then generate a random catalog with the same survey footprint and depth
selection, and finally estimate the angular auto-correlation function together
with its covariance.

\subsection{Quasar sample}

\begin{figure}[htbp]
    \centering
    \includegraphics[width=0.9\textwidth]{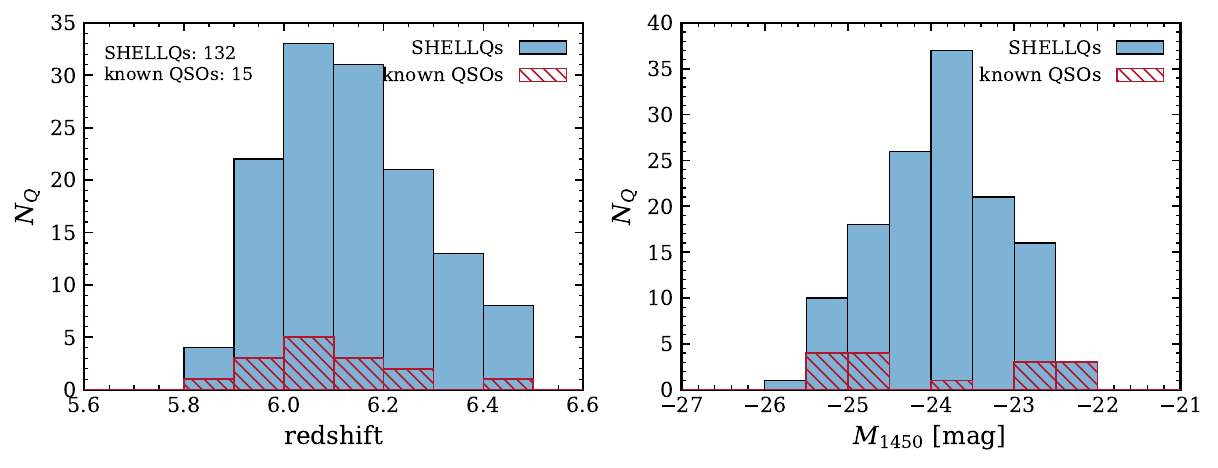}
    \caption{Redshift and UV absolute magnitude distributions of the quasar sample.}
    \label{fig:qso-sample-distribution}
\end{figure}

We adopt the $z\simeq 6$ quasar sample selected from the SHELLQs survey
\cite{Matsuoka_2025,Matsuoka_2021,Matsuoka_2019,Matsuoka_2018,Matsuoka_2017,Matsuoka_2016}
and supplement it with known quasars from the literature
\cite{Banados_2016,Willott_2013,Jiang_2007,Venemans_2015,Willott_2009,Jiang_2009,Mazzucchelli_2017,Kim_2015,Kashikawa_2015,Zeimann_2011}
that lie within the same survey footprint from the HSC-SSP PDR3 Wide-layer dataset
with a sky coverage of $\sim 1200\,{\rm deg}^2$ \cite{Aihara:2021jwb}.
The fiducial sample is restricted to the redshift interval
$5.88 \leq z \leq 6.49$ \cite{Shinohara:2023wjd}, so that the objects probe approximately the
same cosmic epoch as the PBH/SMBH clustering model considered in this work.
Following the reference analysis, the sample spans a broad range of UV absolute
magnitude, approximately $-25.58 \leq M_{1450} \leq -22.25$, and consists
of 132 SHELLQs quasars plus additional 15 known quasars in the same redshift range.
The resulting redshift and UV absolute magnitude distributions of the adopted
sample are shown in Fig.~\ref{fig:qso-sample-distribution}.

\subsection{Random catalog and survey mask}

The estimator of the angular correlation function requires a random catalog that
traces the same angular selection function as the real quasar sample. The
HSC-SSP PDR3 Wide-layer data release provides a random-point catalog with a
surface density of 100 points per arcmin$^2$ for each coadd image and for each
filter \cite{Aihara:2021jwb,Aihara:2019xyr}.
Following the treatment in Ref.~\cite{Shinohara:2023wjd}, we apply to the random
points the same survey masks and quality selections used for the real quasars.
These selections include the photometric-quality flags, the primary-detection
requirement, the input-count threshold, and the bright-star masks specified in
Ref.~\cite{Shinohara:2023wjd}. This procedure ensures that the random catalog
traces the effective survey footprint rather than an idealized uniform sky
coverage.

The spatially varying limiting depth of the survey must also be included,
because non-uniform depth changes the probability of selecting quasars across
the footprint. We therefore follow Ref.~\cite{Shinohara:2023wjd} and assign an
effective depth or magnitude uncertainty to each random position according to
the local imaging properties, then impose the same depth-related selection
threshold as in the real quasar sample. This step allows the random catalog to
reproduce both the angular mask and the inhomogeneous detection efficiency of
the data. Following this procedure, we randomly select 100000 points from the
parent random catalog and apply the above selections. After masking and depth
cuts, 60092 random points remain and are used as the final random quasar catalog
in the angular clustering analysis.

\subsection{Angular auto-correlation estimator}

We estimate the projected two-point angular auto-correlation function using the
Davis--Peebles pair-count estimator \cite{davis1983survey},
\begin{equation}
    \omega(\theta) = \frac{DD(\theta)}{DR(\theta)} - 1 .
\end{equation}
Here $DD(\theta)$ and $DR(\theta)$ are the normalized quasar--quasar and
quasar--random pair counts in an angular bin centered at $\theta$. Explicitly,
\begin{equation}
    DD(\theta) =
    \frac{N_{QQ}(\theta)}{N_Q(N_Q-1)/2},
    \qquad
    DR(\theta) =
    \frac{N_{QR}(\theta)}{N_QN_R},
\end{equation}
where $N_{QQ}$ and $N_{QR}$ denote the numbers of quasar--quasar and
quasar--random pairs in the bin, while $N_Q$ and $N_R$ are the total numbers of
real and random quasars. The pair counts are evaluated in logarithmically spaced
angular bins over the range relevant for the comparison with the PBH clustering
prediction.

The statistical uncertainty is estimated with Jackknife resampling \cite{SDSS:2004oes}. We divide
the survey footprint into $N_{\rm JK}$ subregions, recompute the angular
correlation function after removing one subregion at a time, and construct the
covariance matrix as
\begin{equation}
    C_{ij} =
    \frac{N_{\rm JK}-1}{N_{\rm JK}}
    \sum_{k=1}^{N_{\rm JK}}
    \left[\omega_i^{(k)}-\bar{\omega}_i\right]
    \left[\omega_j^{(k)}-\bar{\omega}_j\right],
\end{equation}
where $\omega_i^{(k)}$ is the correlation function in the $i$-th angular bin for
the $k$-th Jackknife realization, and $\bar{\omega}_i$ is the average over all
realizations. The diagonal elements give the bin-by-bin uncertainties,
$\sigma_i=\sqrt{C_{ii}}$.

\begin{figure}[htbp]
    \centering
    \includegraphics[width=0.9\textwidth]{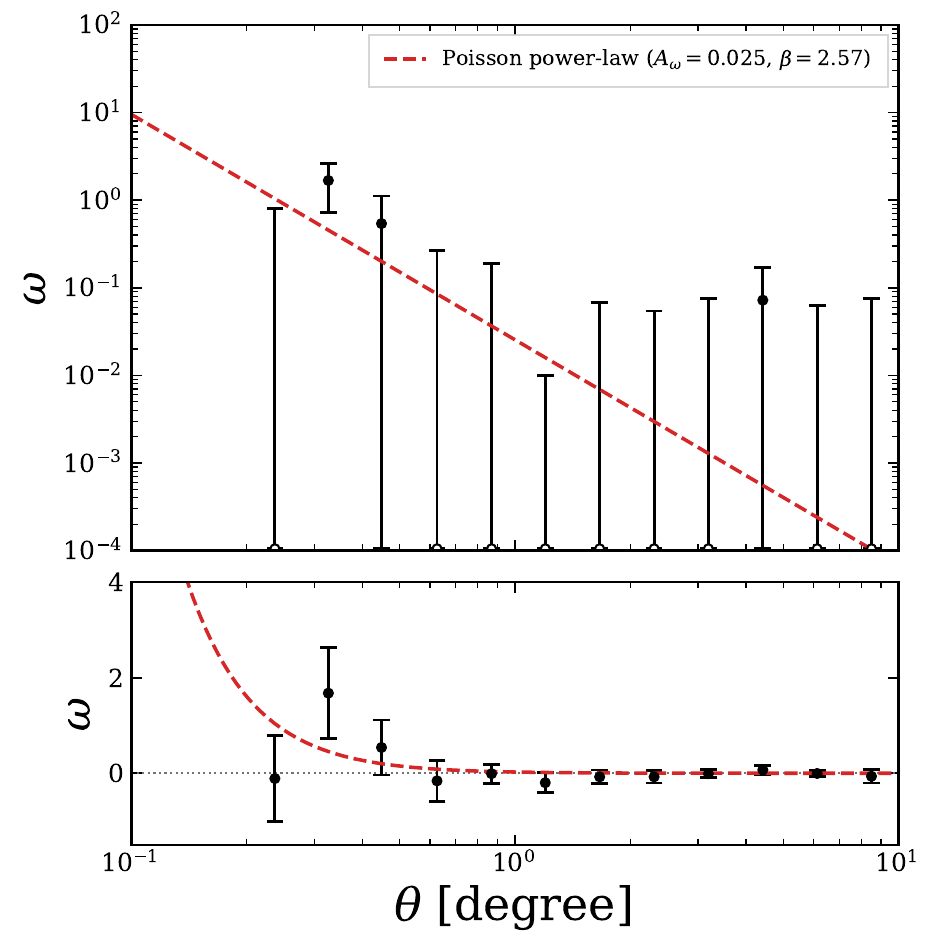}
    \caption{Two-point angular auto-correlation function derived from the
    $z\simeq 6$ quasar sample. The black points show the binned ACF with
    Jackknife uncertainties in logarithmic scale (upper panel) and linear scale
    (lower panel). The red dashed curve denotes the best-fit power-law model
    obtained from the Poisson pair-count likelihood.}
    \label{fig:qso-acf}
\end{figure}

We describe the measured ACF with the power-law form
\begin{equation}
    \label{eq:qso-acf-power-law-fit}
    \omega_{\rm fit}(\theta)
    =
    A_{\omega}
    \left(\frac{\theta}{\rm deg}\right)^{-\beta}.
\end{equation}
The two parameters, $A_{\omega}$ and $\beta$, are determined directly from the
pair counts using a Poisson likelihood, following the strategy of
Ref.~\cite{Croft:1997mv,He:2017crq}. This likelihood-based fit is less sensitive
to negative ACF bins than a standard $\chi^2$ fit, because it is applied to the
pair counts rather than to a prescribed set of binned correlation values.
The resulting ACF, together with the best-fit power-law model, is shown in
Fig.~\ref{fig:qso-acf}; the corresponding pair counts and Jackknife errors are
listed in Table~\ref{tab:qso-acf}. These measurements provide the observational
input for the MCMC comparison with the Poisson and initially clustered PBH
models below.

\begin{table}[htbp]
    \centering
    \caption{Angular auto-correlation function of the quasar sample.}
    \label{tab:qso-acf}
    \begin{tabular}{|c|ccrrcc|}
        \hline
        No. & $\theta\,[\mathrm{deg}]$ & $(\theta-\Delta\theta,\,\theta+\Delta\theta)\,[\mathrm{deg}]$ & $N_{QQ}$ & $N_{QR}$ & $\omega$ & $\sigma_{\mathrm{JK}}$ \\
        \hline\hline
        1 & 0.24 & $(0.2, 0.28)$ & 1 & 925 & -0.110 & 0.907 \\
        2 & 0.33 & $(0.28, 0.38)$ & 5 & 1538 & 1.676 & 0.952 \\
        3 & 0.45 & $(0.38, 0.53)$ & 6 & 3208 & 0.540 & 0.575 \\
        4 & 0.63 & $(0.53, 0.74)$ & 6 & 5874 & -0.159 & 0.427 \\
        5 & 0.87 & $(0.74, 1.02)$ & 13 & 10800 & -0.009 & 0.198 \\
        6 & 1.20 & $(1.02, 1.41)$ & 19 & 19492 & -0.198 & 0.207 \\
        7 & 1.66 & $(1.41, 1.96)$ & 42 & 37408 & -0.076 & 0.144 \\
        8 & 2.30 & $(1.96, 2.71)$ & 73 & 64946 & -0.075 & 0.129 \\
        9 & 3.19 & $(2.71, 3.76)$ & 131 & 108772 & -0.009 & 0.084 \\
        10 & 4.43 & $(3.76, 5.21)$ & 209 & 160475 & 0.072 & 0.097 \\
        11 & 6.13 & $(5.21, 7.22)$ & 253 & 209177 & -0.004 & 0.068 \\
        12 & 8.50 & $(7.22, 10.0)$ & 283 & 248827 & -0.064 & 0.139 \\
        \hline
    \end{tabular}
\end{table}

\section{Results}
\label{sec:results}

We now compare the observed quasar ACF with the angular correlation expected
from PBHs over the same redshift window. Following Ref.~\cite{Shinohara:2021psq},
the angular correlation function of PBHs distributed between
$z_{\rm low}<z<z_{\rm high}$ can be written as a line-of-sight projection,
\begin{equation}
    \label{eq:pbh-angular-projection}
    w_{\rm PBH}(\theta)
    =
    \int_{R_{\rm low}}^{R_{\rm high}} dR_1
    \int_{R_{\rm low}}^{R_{\rm high}} dR_2\,
    \frac{3R_1^2}{R_{\rm high}^3-R_{\rm low}^3}
    \frac{3R_2^2}{R_{\rm high}^3-R_{\rm low}^3}
    \xi_{\rm PBH}(R_1,R_2,\theta).
\end{equation}
Here $R_{\rm low}$ and $R_{\rm high}$ are the comoving distances corresponding
to the lower and upper redshift cuts of the quasar sample. In the numerical
analysis we set $z_{\rm low}=5.88$ and $z_{\rm high}=6.49$, matching the
redshift range of the quasar catalog, and evaluate the double integral on a
uniform grid in comoving distance.
The redshift-dependent evolution factors in both the Poisson and initially
clustered models are evaluated at the effective survey redshift
$z_{\rm mean}=\frac{1}{2}(z_{\rm low}+z_{\rm high})$.
Across this redshift interval, the evolution factors evaluated at
$z_{\rm low}$ and $z_{\rm high}$ differ only mildly. We therefore use the
evolution at $z_{\rm mean}$ as a representative approximation for the whole
quasar sample.

For each angular bin we compare the observed quasar pair counts with the model
prediction. Given a model angular correlation function $w_{\rm model}(\theta)$,
the expected number of quasar--quasar pairs in the $i$-th bin is written as
\begin{equation}
    h_i =
    \left[1+w_{\rm model}(\theta_i)\right]
    DR_i,
\end{equation}
Following Refs.~\cite{Shinohara:2023wjd,Croft:1997mv,He:2017crq}, we use the
Poisson pair-count likelihood
\begin{equation}
    \label{eq:mcmc-pair-count-likelihood}
    \mathcal{L}
    =
    \prod_i
    \frac{
    e^{-h_i} h_i^{N_{QQ,i}}
    }{
    N_{QQ,i}!
    },
\end{equation}
where $N_{QQ,i}$ is the observed number of quasar--quasar pairs in the $i$-th
bin. We minimize
\begin{equation}
    \label{eq:mcmc-pair-count-statistic}
    S
    =
    -2\ln\mathcal{L}
    =
    2\sum_i
    \left(
    h_i
    -
    N_{QQ,i}\ln h_i
    \right),
\end{equation}
where the factorial term is parameter independent. This is the same likelihood
structure used for fitting the observed ACF, but now with $w_{\rm model}$
supplied by the PBH correlation model.

We consider two model classes. For the Poisson initial condition model, the free
parameters are
\begin{equation}
    \left(\log_{10}\fpbh,\log_{10}\mpbh\right),
\end{equation}
The real-space correlation function is taken from
Eq.~\eqref{eq:poisson-correlation-evolution}\footnote{In the numerical
projection, the large-scale linear branch is continued with the quasi-linear
expression rather than set exactly to zero, so that the piecewise correlation
function remains continuous at $x=x_{L-QL}(z)$.} and inserted into the
projection formula \eqref{eq:pbh-angular-projection}.

For the initially clustered model, the angular projection is mainly sensitive to
the clustering amplitude at the redshift probed by the quasar sample. It is
therefore difficult for $w(\theta)$ alone to distinguish the initial amplitude
$\xi_0$ from the subsequent contraction parameter $\lambda$. We absorb this
degeneracy into an effective correlation amplitude,
\begin{equation}
    \xi_{\rm eff}
    =
    (1+\xi_0)
    \left(\frac{1+z_{\rm eq}}{1+z_{\rm mean}}\right)^{3\lambda}
    -1,
\end{equation}
The MCMC analysis is then performed with the reduced parameter vector
\begin{equation}
    \left(\xi_{\rm eff},\rcl\right),
\end{equation}
This parametrization keeps the observable amplitude and
scale explicit, while avoiding an artificial separation of $\xi_0$ and
$\lambda$ that is not resolved by the present angular correlation function.

\begin{figure}
    \centering
    \includegraphics[width=0.75\textwidth]{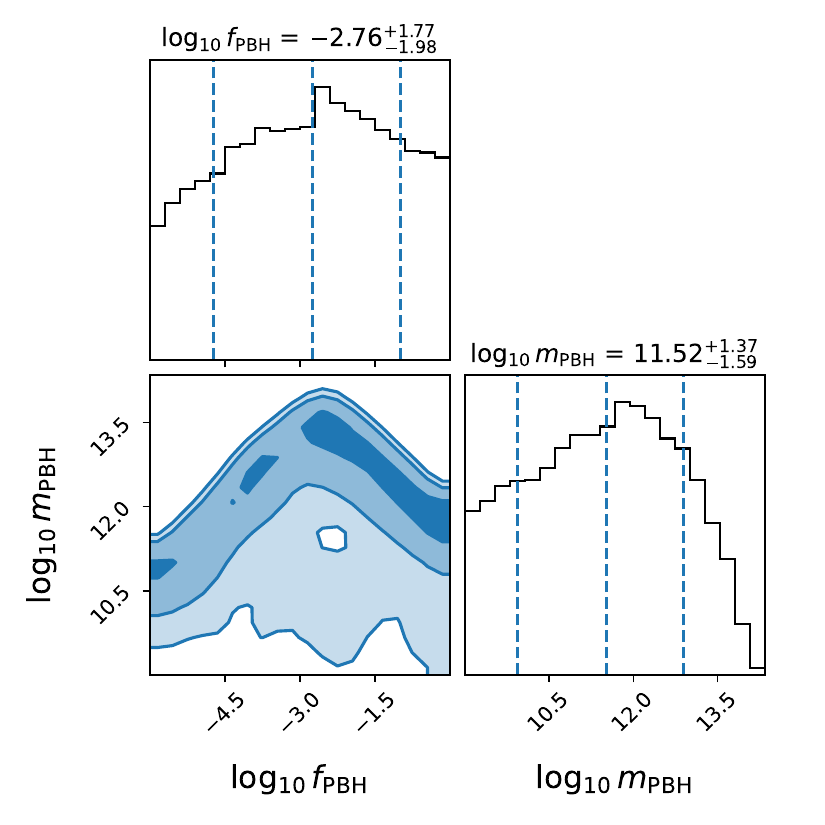}
    \caption{Posterior distribution for the Poisson initial condition model.
    The diagonal panels show the marginalized constraints on the sampled
    parameters, while the off-diagonal panel shows their joint posterior
    distribution inferred from the quasar angular correlation function.}
    \label{fig:corner-poisson}
\end{figure}

The resulting posterior constraints for the Poisson model are shown in
Fig.~\ref{fig:corner-poisson}. The posterior favors a small PBH abundance, with
the preferred region around $\fpbh\sim 10^{-3}$, while the PBH mass is driven
toward the supermassive range, $\mpbh\sim 10^{12}M_\odot$. In the present model,
however, $\mpbh$ should not be identified directly with the mass of the central
black hole powering an individual quasar. Instead, the PBH mass controls the
Poisson initial fluctuations and the subsequent formation and clustering
properties of the quasar host halos. The angular correlation function therefore
mainly selects an effective mass scale associated with the host-halo population,
rather than the central SMBH mass itself. In this sense, the preferred value of
$\mpbh$ lies in the range broadly compatible with the massive host environments
inferred from quasar clustering measurements
\cite{Shen:2007xk,Eftekharzadeh:2015xoa,Arita:2023xsp}.

\begin{figure}
    \centering
    \includegraphics[width=0.75\textwidth]{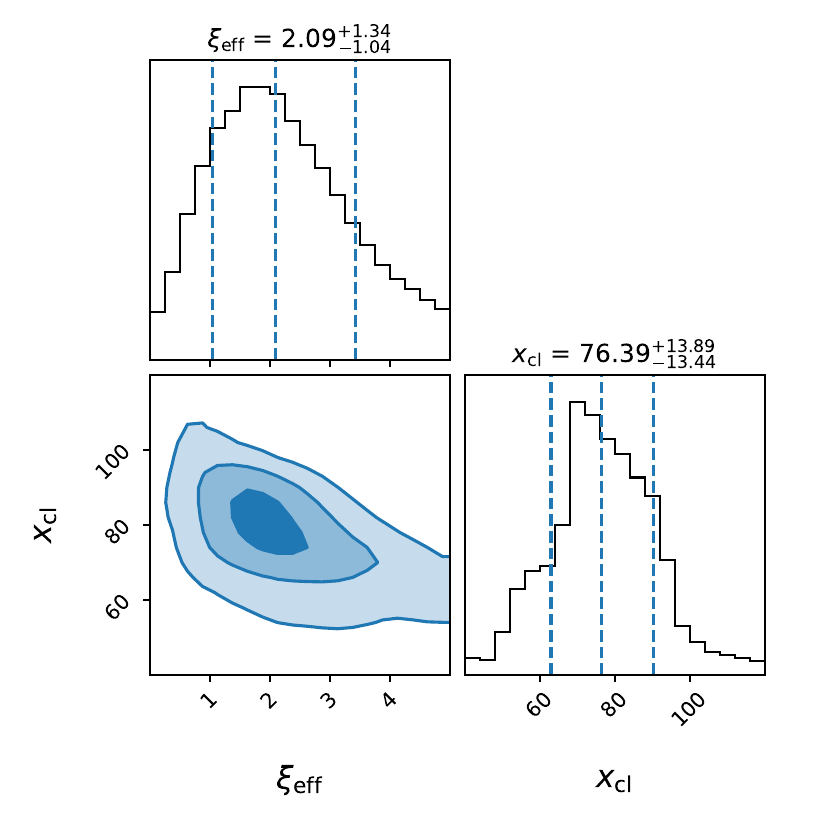}
    \caption{Posterior distribution for the initially clustered model written in
    terms of the effective clustering amplitude $\xi_{\rm eff}$ and the cluster
    scale $\rcl$. The parameters describe the effective amplitude and boundary
    scale of the top-hat cluster.}
    \label{fig:corner-cluster}
\end{figure}

Fig.~\ref{fig:corner-cluster} presents the corresponding result for the
initially clustered model. In this parametrization, the angular correlation
function favors an order-unity effective clustering amplitude,
$\xi_{\rm eff}\simeq 2.1$, together with a characteristic cluster scale
$\rcl\simeq 76\,{\rm Mpc}$ in the comoving units used in the fit. Since the
clustered profile is modeled as a spherical top-hat, these two parameters have
direct physical meanings: $\xi_{\rm eff}$ fixes the excess pair probability
inside the cluster, while $\rcl$ marks the boundary of the top-hat cluster. The
current angular correlation measurements therefore constrain both the amplitude
and the characteristic scale of the clustered profile.

\begin{figure}
    \centering
    \begin{subfigure}{0.48\textwidth}
        \centering
        \includegraphics[width=\textwidth]{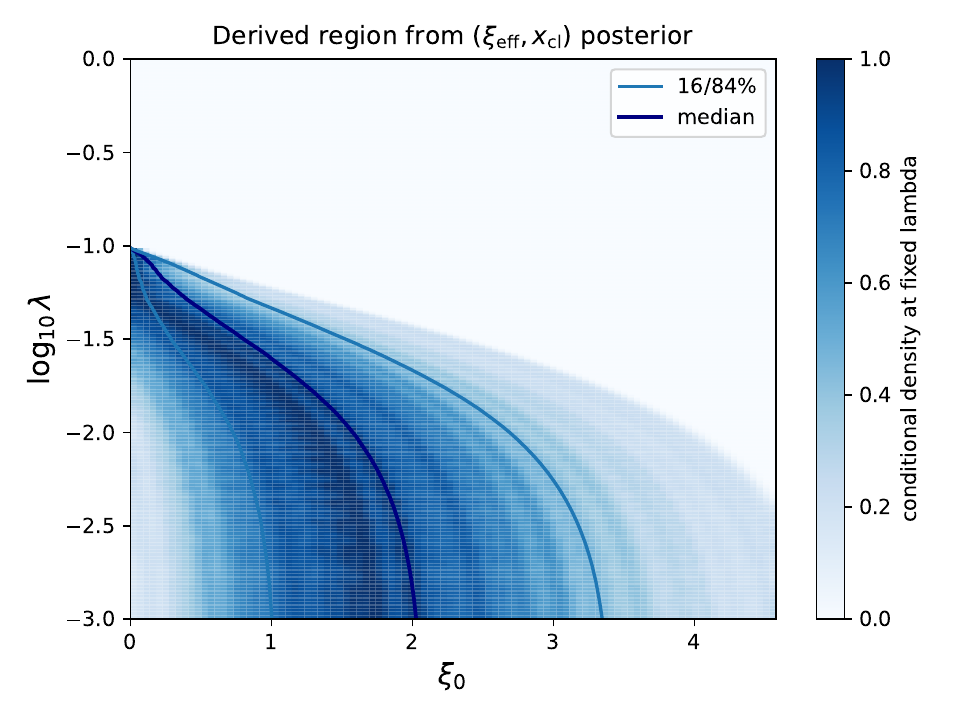}
        \caption{$(\xi_0,\log\lambda)$}
        \label{fig:derived-xi0-loglambda}
    \end{subfigure}
    \hfill
    \begin{subfigure}{0.48\textwidth}
        \centering
        \includegraphics[width=\textwidth]{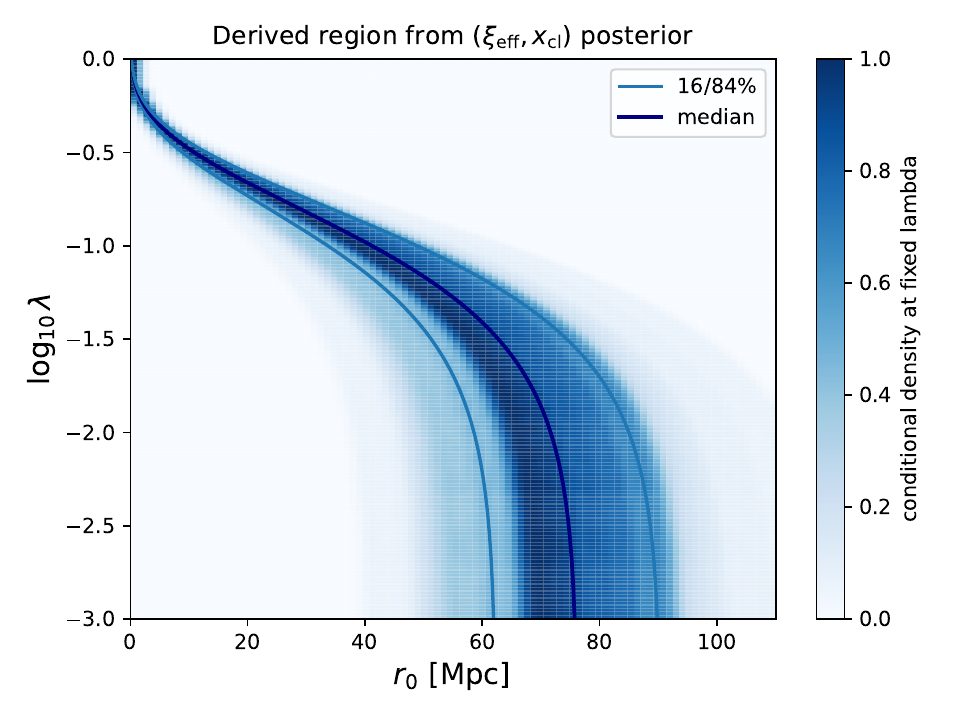}
        \caption{$(r_0,\log\lambda)$}
        \label{fig:derived-r0-loglambda}
    \end{subfigure}
    \caption{Derived conditional distributions for the original parameters of
    the initially clustered model. The left panel maps the effective clustering
    constraint back to the initial amplitude $\xi_0$ and the evolution parameter
    $\lambda$, while the right panel translates the inferred effective cluster
    scale into the initial comoving radius $r_0$ for fixed $\lambda$.}
    \label{fig:derived-cluster-parameters}
\end{figure}

To connect the effective parametrization back to the original cluster variables,
we also reconstruct conditional distributions for $(\xi_0,\lambda)$ and
$(r_0,\lambda)$ using the relation between $\xi_{\rm eff}$ and the evolved
cluster correlation amplitude. These derived constraints are shown in
Fig.~\ref{fig:derived-cluster-parameters}. The reconstruction indicates that the
same effective clustering signal can be produced by different combinations of
initial amplitude and contraction rate. Requiring a positive initial clustering
amplitude, $\xi_0>0$, restricts the contraction parameter to
$\lambda\lesssim 10^{-1}$, suggesting that the relative contraction of PBH
pairs in comoving coordinates is relatively weak. The inferred $\xi_0$ and
$r_0$ vary rapidly in the range $10^{-2}\lesssim\lambda\lesssim 10^{-1}$,
whereas for $\lambda\lesssim 10^{-2}$ the constraints become nearly insensitive
to $\lambda$. We expect $\lambda$ to depend on the PBH mass and abundance, with
larger $\mpbh$ and $\fpbh$ leading to stronger contraction, but a quantitative
model for this dependence is left for future work.

\section{Conclusions}
\label{sec:conclusions}

In this work we have investigated the angular auto-correlation function of
high-redshift quasars as a probe of supermassive PBH clustering. We considered
two classes of initial conditions: the Poisson fluctuations associated with the
discreteness of PBHs, and an initially clustered PBH population described by a
spherical top-hat correlation profile \cite{Luca_2020,DeLuca_2022,Huang_2024}.
In both cases, the three-dimensional PBH
correlation function was evolved to the quasar epoch and projected over the
observed redshift window, allowing a direct comparison with the measured quasar
angular correlation function \cite{Shinohara:2021psq,Shinohara:2023wjd}.

Within the Poisson initial condition model, the posterior prefers a small PBH
abundance, $\fpbh\sim 10^{-3}$, and a PBH mass in the supermassive range,
$\mpbh\sim 10^{12}M_\odot$. In the interpretation adopted here, this mass should
be understood as an effective scale controlling the formation and clustering of
quasar host halos, rather than as the mass of the central black hole in an
individual quasar \cite{Shen:2007xk,Eftekharzadeh:2015xoa,Arita:2023xsp}.
For the initially clustered model, the angular correlation
function favors an effective clustering amplitude $\xi_{\rm eff}\simeq 2.1$ and
a top-hat cluster boundary scale $\rcl\simeq 76\,{\rm Mpc}$. By mapping these
effective parameters back to the original variables, we further obtained
conditional constraints on $\xi_0$ and $r_0$. The requirement $\xi_0>0$ suggests
$\lambda\lesssim 10^{-1}$, indicating that the relative contraction of PBH pairs
in comoving coordinates is weak within the range allowed by the present data.

There are several directions in which this analysis can be extended. First, a more complete description should
allow astrophysical black holes and PBHs to coexist, so that their separate and
combined contributions to the angular correlation function can be modeled
consistently \cite{Harikane:2015unm}. Second, improved measurements of quasar clustering on smaller
angular scales would be particularly valuable, since the present sample contains
only a limited number of close quasar pairs and the corresponding uncertainties
remain large. Finally, the theoretical model can be refined by including PBH
mergers \cite{Huang:2023klk,Hai-LongHuang:2023atg}, which may affect the cluster evolution through merger-driven mass
growth, changes in the PBH number density, and modifications of the clustering
profile.

\section*{Acknowledgments}

This work is supported by NSFC, No.12475064, National Key Research
and Development Program of China, No. 2021YFC2203004, and the
Fundamental Research Funds for the Central Universities.

\bibliography{REF}

\end{document}